\documentclass[sigconf]{acmart}

\usepackage{enumitem}

\usepackage{tikz}
\usepackage{array}

\usetikzlibrary{external, shapes.geometric, arrows}

\AtBeginDocument{%
  \providecommand\BibTeX{{%
    \normalfont B\kern-0.5em{\scshape i\kern-0.25em b}\kern-0.8em\TeX}}}

\begin{document}

\tikzstyle{5_box_node} = [
    rectangle,
    rounded corners, 
    minimum width=2cm, 
    minimum height=1cm,
    text centered,
    text width = 2.5cm,
    draw=black,
]
\tikzstyle{3_box_node} = [
    rectangle,
    rounded corners, 
    minimum width=3cm, 
    minimum height=1cm,
    text centered,
    text width = 4cm,
    draw=black,
]
\tikzstyle{4_box_node} = [
    rectangle,
    rounded corners, 
    minimum width=3cm, 
    minimum height=1cm,
    text centered,
    text width = 3.2cm,
    draw=black,
]
\tikzstyle{arrow} = [thick,->,>=stealth]


\newcommand{\yaqing}[1]{\textcolor{violet}{#1}}
\newcommand{\vm}[1]{\textcolor{blue}{#1}}
\newcommand{\sssec}[1]{\vspace*{0.05in}\noindent\textbf{#1}}
\newcommand{\modelname}{knowledge space model}

\newcommand{\revision}[1]{\textcolor{black}{#1}}

\title{From Overload to Insight: Scaffolding Creative Ideation through Structuring Inspiration}

\author{Yaqing Yang}
\affiliation{%
  \institution{Carnegie Mellon University}
  \city{Pittsburgh, PA}
  \country{USA}}
\email{yaqingyy@cs.cmu.edu}

\author{Vikram Mohanty}
\affiliation{%
  \institution{Carnegie Mellon University}
  \city{Pittsburgh, PA}
  \country{USA}}
\email{vikrammohanty@acm.org}

\author{Nikolas Martelaro}
\affiliation{%
  \institution{Carnegie Mellon University}
  \city{Pittsburgh, PA}
  \country{USA}}
\email{nikmart@cmu.edu}

\author{Aniket Kittur}
\affiliation{%
  \institution{Carnegie Mellon University}
  \city{Pittsburgh, PA}
  \country{USA}}
\email{nkittur@cs.cmu.edu}

\author{Yan-Ying Chen}
\affiliation{%
  \institution{Toyota Research Institute}
  \city{Los Altos, CA}
  \country{USA}}
\email{yan-ying.chen@tri.global}

\author{Matthew K. Hong}
\affiliation{%
  \institution{Toyota Research Institute}
  \city{Los Altos, CA}
  \country{USA}}
\email{matt.hong@tri.global}

\renewcommand{\shortauthors}{Yang et al.}

\begin{abstract}

 Creative ideation relies on exploring diverse stimuli, but the overwhelming abundance of information often makes it difficult to identify valuable insights or reach the `aha' moment. Traditional methods for accessing design stimuli lack organization and fail to support users in discovering promising opportunities within large idea spaces. In this position paper, we explore how AI can be leveraged to structure, organize, and surface relevant stimuli, guiding users in both exploring idea spaces and mapping insights back to their design challenges. 


\end{abstract}



\settopmatter{printacmref=false}
\setcopyright{none}
\renewcommand\footnotetextcopyrightpermission[1]{}
\pagestyle{plain}
\maketitle

\section{Introduction}
\label{intro} 
A key driver in innovative problem solving, such as in design or scientific discovery, is exposure to a broad and diverse set of inspirations that can motivate potential new solutions ~\cite{dow2010parallel}. These inspirations serve as external stimuli that help in problem formulation, direction definition, and solution generation ~\cite{gonccalves2016inspiration}.
Today people have access to more sources of inspiration than ever before, using online search engines and generative AI to find or create relevant solution mechanisms from nearly any source, whether a web page, video, patent, or paper. However, finding, and making sense of, and using this enormous space of inspirations is limited because of how challenging it is for users to browse through, organize, and select specific ideas~\cite{toner2024artificial, xu2024idea}. Though the stimuli retrieved or generated by computational system have the potential to spark the a-ha! moment during ideation~\cite{koch2019may,kwon2023understanding,xu2024idea, gonccalves2016inspiration,ma2023conceptual}, users can get overwhelmed in this large volume of external information. 

Here we explore two core challenges that are likely to be critical in helping harness the power of both retrieval- and generation-powered systems for inspiring designers as the number of ideas they have access to continues to increase and become overwhelming. First, we investigate ways of structuring those ideas so that people can more efficiently explore them, focusing on areas of potential interest and avoiding areas they think may be less relevant or fruitful. In order to support this we introduce a new lens on organization: instead of topic-based approaches we explore computationally extracting and organizing the functional mechanisms of inspirations (stimuli) to create `mechanism trees'. While mechanism trees have been used in the past~\cite{linsey2012design, shah2003metrics}, concretely for measuring exploration of a design space or more inspirationally for structuring that exploration, most approaches have to date relied on manual creation and navigation.

Second, we explore the challenge of helping people notice and transfer the `active ingredient' of an inspiration, which might be from a very different domain, to their problem domain. For example, an ultrasonic cleaning system developed for removing contaminants from spacecraft might be useful for a designer interested in reducing the amount of water used in laundry, but might be difficult for them to initially notice how it might be relevant as well as how to actually transfer and use it for the laundry domain. To probe this we use large language models (LLMs) to generate inspirations at different levels of distance from the original problem domain and examine what kinds of information people use to connect to and use those inspirations.

\section{Prototype Design}

\subsection{Structuring Design Stimuli through Functional Hierarchies}
Many studies have shown that structured data representation efficiently supports users to make sense of AI-generated content ~\cite{gero2024supporting, jiang2023graphologue}. To support design space exploration, studies have organized and presented design materials based on their similarity using clusters, maps, charts and hierarchical structures~\cite{suh2024luminate, siangliulue2016ideahound, chung2024patchview, dove2016argument, biskjaer2014constraint,linsey2012design, chen2024autospark}. These organizations  have shown promising results with categorization for efficiently making sense of external inspirations.
Among those structures, the hierarchical structure provides multiple levels of concept abstraction and offers a clear, organized view. For instance, the WordTree method \cite{linsey2012design} arranges design concepts by placing general (hypernym) ideas at higher levels and more specific (troponym) ideas at lower levels. This supports broader thinking by enabling exploration of concepts at varying levels of granularity. Leveraging such a structure allows people to navigate between broad topics and more detailed stimuli for deeper exploration.  

We hypothesize that categorizing stimuli, i.e. the extracted mechanisms of the inspirations, by  functional similarity~\cite{hope2022scaling} within a hierarchical structure will help users  scan high-level topics and easily identify novel, non-obvious stimuli, guiding them to efficiently decide where to invest further effort. Once they find intriguing directions, they can dive into lower-level mechanisms at different degrees of granularity to explore potential solutions in depth. However, prior work like WordTree~\cite{linsey2012design} was largely relied on manual work which is hard to scale and though there are systems like AI mind-map tools \footnote{https://monica.im/tools/ai-mind-map-maker} that generate ideas in hierarchical formats, there's no prior work explored how to automatically organize large scale of given unstructured stimuli into hierarchical structure to support solution space exploration. 

\begin{figure}[h]
  \centering
  \includegraphics[width=\linewidth]{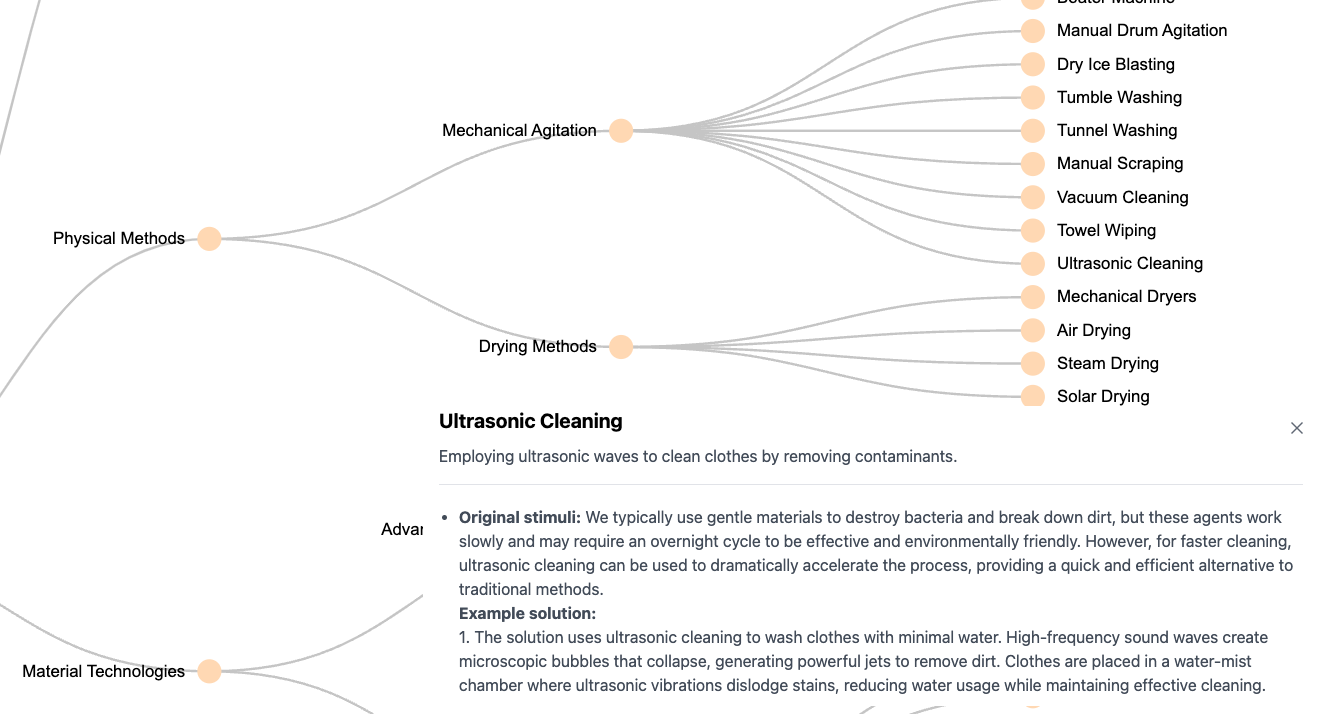}
  \caption{Mechanism Tree for a design task: laundry with less water}
  \label{fig:tree}
\end{figure}

To address the gap, we built a prototype system that automatically organize stimuli from an unstructured format into a hierarchical tree structure using a LLM pipeline based on ~\cite{wan2024tnt}. Those unstructured stimuli could be any textual descriptions of possible solutions to a problem, like those solutions collected from crowd-sourcing or generated by LLM. The system will automatically generate a mechanism taxonomy to categorize the unstructured stimuli. 
Figure \ref{fig:tree} demonstrates the mechanism tree for solving the problem `clean laundry with less water.'  Abstract concepts appear at higher levels, and more concrete concepts are placed lower. The bottom card provides a description of the mechanism, the original stimuli and a sample solution demonstrating its use.

\subsection{Providing Analogical Cues for Cross-Domain Adaptation}
When people have only vague ideas of the solution, such as just having encountered stimuli of interests during ideation, they are likely to benefit from divergent thinking, which can be supported by  providing them with not only relevant but also semantically diverse stimuli ~\cite{xu2024idea}. Analogical thinking can effectively uncover shared structural similarities of semantically diverse stimuli~\cite{linsey2012design,gick1983schema}, thus might useful for supporting further exploration of stimuli of interests. Analogical thinking requires people to transfer the functional mechanism or `active ingredients' that solve a problem in one domain to another domain where it may also solve the problem~\cite{hope2022scaling,goucher2018inspired,hope2017accelerating,linsey2012design}. However, moving these ingredients from source to target domains demands significant effort and expertise~\cite{gentner1983structure,chan2011benefits}, so automating this process using AI could greatly assist practitioners. Although researchers have automated the analogical search for stimuli and extracted active ingredients from diverse fields~\cite{gilon2018analogy,hope2022scaling,kang2022augmenting},computational methods for effectively supporting the adaptation of these stimuli and ingredients remain underexplored.

\begin{figure}[h]
  \centering
  \includegraphics[width=\linewidth]{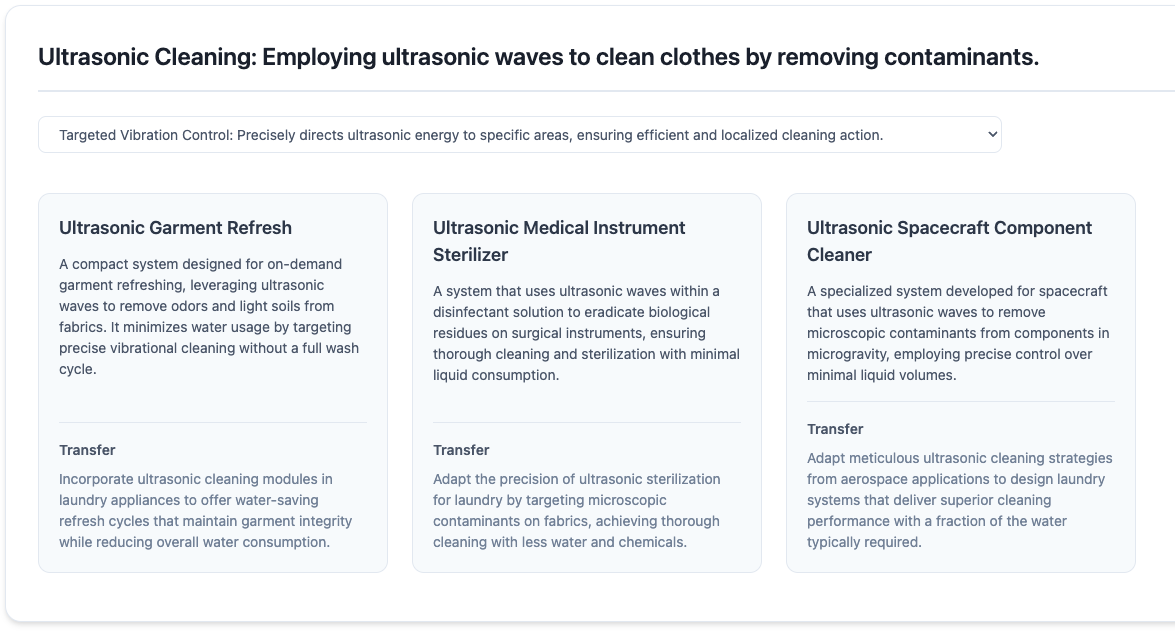}
  \caption{Examples for the mechanism `ultrasonic cleaning' for solving the problem: laundry with less water}
  \label{fig:example}
\end{figure}

We hypothesize that providing users with certain analogical thinking cues might be useful for supporting their analogical thinking, so they could creatively adapt the stimuli of interest to their original problem. 
Prior research shows that solution examples from near or far domains offer distinct benefits for problem-solving~\cite{fu2013meaning,chan2011benefits}, suggesting the most helpful examples often come from a `sweet spot' between near and far~\cite{fu2013meaning}. Prior research has also shown that expert rating of example innovation potential correlates weakly with domain distance~\cite{goucher2020adaptive}.
This potential stems from each example’s active ingredients and their feasibility for adaptation. Based on these findings, we generated three types of cues using LLM for each stimuli in the tree structure (Figure ~\ref{fig:tree}) and incorporated them in our prototype system: 

\begin{enumerate}
    \item \textbf{{Diverse Cross-Domain Applications of Stimuli}.} For each stimulus in the tree view (Figure ~\ref{fig:tree}), we generated \textit{close}, \textit{somewhat far}, and \textit{distant} application examples relative to the original problem domain to provide relevant yet semantically diverse information for supporting diverse thinking. Figure ~\ref{fig:example} illustrates three examples applying the stimulus `ultrasonic cleaning' from Figure ~\ref{fig:tree}. These examples come from domains that are close (garment refresh), somewhat far (medical instruments), and distant (spacecraft) relative to the original domain (laundry).
    \item \textbf{Core Functional Mechanisms (`Active Ingredients').} For each example, we used LLM to extract its active ingredients for implementing the stimuli and use the ingredients as filters, allowing users to view examples with same active ingredients across domains. As shown in Figure~\ref{fig:example}, the three examples come from different domains but share the principle of `target vibration control' when adopting `ultrasonic cleaning.' 
    \item \textbf{Actionable Transfer Strategies for Adaptation.} We also generate a brief transfer demonstration at the bottom of each example card, illustrating potential stimuli adaptation inspired by the example and the active ingredients. For instance, a `medical instrument sterilizer' might suggest a transfer like: \emph{`Adapt the precision of ultrasonic sterilization for laundry by targeting microscopic contaminants on fabrics, achieving thorough cleaning with less water and chemicals.'}
\end{enumerate}

\section{Preliminary Findings}

We conducted a pilot study with four researchers from robotics, fabrication, and sensing. Each participant proposed a problem to brainstorm in their field. We then generated 100 potential mechanisms and organized them into a hierarchical structure using GPT-4o. Participants compared a mechanism list view with a function-based hierarchical view (see Figure~\ref{fig:tree}). We followed up with a detailed case study involving a researcher in computer vision, repeating these steps. For three mechanisms the participant found interesting, we provided diverse  examples with active ingredients and transfer in the same format as shown in  Figure ~\ref{fig:example}. All the data for the detailed case study is generated using GPT-o1-mini. We interviewed all participants after they used the prototype system, asking about the types of stimuli they found interesting and their experience in discovering and leveraging them for ideation using our system.

\subsection{Stimuli of Interest} 
 Participants identified three categories of stimuli they found valuable.

\subsubsection{\textbf{Familiar yet Feasible Ideas}}

All participants encountered ideas similar to known feasible solutions within their domain of expertise. While these ideas were not particularly novel, they enhanced trust in the system. As one participant described, they are \emph{essential considerations} for their problem and should be included. This observation aligns with prior research ~\cite{gonccalves2016inspiration}, which suggests that designers often select stimuli based on their perceived relevance to the problem at hand. 

\subsubsection{\textbf{Previously Overlooked but Relevant Concepts}}

Some stimuli triggered recognition of concepts that participants already knew but had not yet considered for their specific problem. These insights helped reframe their thinking, offering alternative approaches they might have otherwise overlooked. For instance, a researcher (P1) working on a 4D editing problem in computer vision noted, \emph{"Though Fourier Transform is something basic, we haven't tried this in the task... it provides another way of decomposing and compression. Rather than using a motion basis, it reminds me that I can use a frequency-based method to decompose the object motion."}

\subsubsection{\textbf{Novel or Unexpected Ideas}}

Participants also encountered unfamiliar concepts that, while not immediately applicable, sparked curiosity and potential for new solutions. For instance, one participant (P2), who was looking for ways to add tank movement to a robotic platform, noticed a stimulus called \textit{Air Cushion Tracks} and remarked, \textit{``Some of these are very cool, like air cushion tracks. This is not what I had...I would not have thought of this idea if I did not look at this tree.''} Even when these stimuli could not be directly implemented, they encouraged creative thinking and exploration of related problem spaces. Some participants noted that these ideas were particularly useful for adjacent tasks rather than the specific problem they were brainstorming. As P2 noted, \textit{`` I'm only making the robotic platform to complement my software system, so probably want something simple...but if my research was focused on the robotic platform itself, I definitely want to explore some of the suggestions here, like air cushion tracks.''}


\subsection{Hierarchical Structure and Analogical Cues}

\subsubsection{\textbf{Hierarchical Structure for Efficient Exploration}}

Participants found the hierarchical (tree) view more effective for discovering relevant stimuli compared to a simple list view. While both views contained the same mechanisms, the tree structure better aligned with the participants’ expectations and facilitated the discovery of novel ideas. In contrast, with the list view, they primarily noticed familiar mechanisms that they had already deemed useful. By organizing stimuli into high-level categories, the hierarchy allowed participants to skip irrelevant branches and focus on promising directions. 

Additionally, stimuli with similar functional meanings were grouped under the same high-level concepts, making it easier for users to recognize connections between ideas. Participants noted that they could "\emph{go beyond the keywords to find more relevant things}" at different levels of abstraction, whereas the list view often limited their exploration to the most immediately recognizable terms.

\subsubsection{\textbf{Analogical Cues for Creative Adaptation}}

When participants explored specific stimuli in depth, they found analogical cues—such as diverse application examples and extracted functional components—helpful for adapting and repurposing ideas. These cues provided concrete references for how similar mechanisms had been applied across different domains, inspiring new ways to integrate them into their own problem spaces.

For example, while examining application examples of the stimulus \emph{"Soft Skeleton Embedding,"} a participant working on 4D editing in computer vision noticed how one example addressed sensor noise. They remarked, \emph{"This could solve the current noisy data problem and possibly even be added on top of other pipelines."} However, participants also noted that the static nature of these examples did not always provide enough guidance for fully adapting a stimulus (and its active ingredients) to their specific problem. 



\section{Conclusion}
\label{sec:discussion}
People browse large amounts of information to generate innovative solutions. We seek design insights for `tools for thoughts' that facilitate efficient solution space exploration for problem-solving. From a preliminary study, we identified three types of stimuli people are interested in when they browse information provided by computational system during ideation process. We hypothesized that 1) the hierarchical structure can help users in quickly skim large volumes of stimuli, identify stimuli of interest for solving their problems, and explore similar concepts of this stimuli at different abstraction levels; 2) providing analogical cues for these stimuli can support the creative adaptation of these stimuli. We incorporated those hypothesis into a prototype system leveraging AI and the pilot study results align with our expectations.

However, our preliminary findings also raised questions. For example, as people find the static transfer might not be helpful, how can we better demonstrate the transfer of the stimuli and the active ingredients in its application examples to support more efficient adaptation?  One method could be allowing users to iteratively interact with an LLM agent to refine a given solution transfer to the original problem.
Future work could also analyze how organizing different examples for a given stimuli, such as according to analogical distance to the original problem, or according to different active ingredients –influences the discovery of novel, creative solutions for the original problem. 
Finally, our prototype solution sets are also small compared to the vast resources available that can be obtained. How can we refine visualizations and interactions to help users navigate large-scale data more efficiently? By addressing these challenges, we can move closer to building more effective AI-powered ideation tools that seamlessly integrate into the creative problem-solving process.




\bibliographystyle{ACM-Reference-Format}
\bibliography{main}

\end{document}